# Thermodynamic and kinetic modeling of Mn-Ni-Si precipitates in low-Cu reactor pressure vessel steels[1]

Huibin Ke [a], Peter Wells [b], Philip D. Edmondson [c], Nathan Almirall [d], Leland Barnard [e], G. Robert Odette [b, d], Dane Morgan [a, f, *]

[a] Department of Materials Science and Engineering, University of Wisconsin, Madison, WI 53706, USA

[b] Mechanical Engineering Department, University of California, Santa Barbara, CA 93106, USA

[c] Materials Science and Technology Division, Oak Ridge National Laboratory, Oak Ridge, TN 37831, USA

[d] Materials Department, University of California, Santa Barbara, CA 93106, USA

[e] Elysium Industries, Boston, MA 02111, USA

[f] Materials Science Program, University of Wisconsin, Madison, WI 53706, USA

[*] Corresponding author. E-mail address: ddmorgan@wisc.edu

## Abstract

Formation of large volume fractions of Mn-Ni-Si precipitates (MNSPs) causes excess irradiation embrittlement of reactor pressure vessel (RPV) steels at high, extended-life fluences. Thus, a new and unique, semi-empirical cluster dynamics model was developed to study the evolution of MNSPs in low-Cu RPV steels. The model is based on CALPHAD thermodynamics and radiation enhanced diffusion kinetics. The thermodynamics dictates the compositional and temperature dependence of the free energy reductions that drive precipitation. The model treats both homogeneous and heterogeneous nucleation, where the latter occurs on cascade damage, like dislocation loops. The model has only four adjustable parameters that were fit to an atom probe tomography (APT) database. The model predictions are in semi-quantitative agreement with systematic Mn, Ni and Si composition variations in alloys characterized by APT, including a sensitivity to local tip-to-tip variations even in the same steel. The model predicts that heterogeneous nucleation plays a critical role in MNSP formation in lower alloy Ni contents. Single variable assessments of compositional effects show that Ni plays a dominant role, while even small variations in irradiation temperature can have a large effect on the MNSP evolution. Within typical RPV steel ranges, Mn and Si have smaller effects. The delayed but then rapid growth of MNSPs to large volume fractions at high fluence is well predicted by the model. For purposes of illustration, the effect of MNSPs on transition temperature shifts are presented based on well-established microstructure-property and property-property models.

---

[1] Notice of Copyright This manuscript has been authored by UT-Battelle, LLC under Contract No. DE-AC05-00OR22725 with the U.S. Department of Energy. The United States Government retains and the publisher, by accepting the article for publication, acknowledges that the United States Government retains a non-exclusive, paid-up, irrevocable, world-wide license to publish or reproduce the published form of this manuscript, or allow others to do so, for United States Government purposes. The Department of Energy will provide public access to these results of federally sponsored research in accordance with the DOE Public Access Plan (http://energy.gov/downloads/doe-public-access-plan).

## 1. Introduction and Background

Nuclear power contributes about 19.5% of the electricity supply in the United States [1], and about 13% worldwide [2]. If this major source of carbon-free energy is to be sustained, life extension of the current fleet of light water reactors will be required to bridge the gap to new plant builds [3]. Life extension will require clear demonstration of large safety margins and reliable and economic long-term plant operation. A key challenge is understanding and managing a large number of materials aging and degradation issues, ranging from concrete and cables, to reactor internals, to pressure boundary steels [4, 5]. Thus, extensive aging research is being conducted to support extending plant life from 40 (the original license) and 60 years (the first license extension), and now to 80 years, or more, based on a second license extension.

One critical life extension issue is neutron irradiation embrittlement of reactor pressure vessels (RPVs). RPVs are massive, thick-walled, permanent structures, whose primary function is to pressurize water to 7 (Boiling Water Reactors, BWRs) to 14 MPa (Pressurized Water Reactors, PWRs), thereby permitting reactor operating temperatures around 290°C [6]. RPVs are also an important barrier to the release of radioactivity in the event of a core damaging accident. Regulations require very low RPV failure probabilities by crack propagation, both for normal operation and postulated low probability accident events. In the unirradiated condition, low alloy RPV steels are very tough, and vessel fracture probabilities are negligibly small, representing no significant risk. However, neutrons leaking from the reactor core cause irradiation hardening and embrittlement, manifested as upward shifts in the ductile-to-brittle transition temperature, $\Delta T$, that may challenge continued vessel operation for some plants during extended life. The sources of embrittlement include formation of precipitates, stable and unstable matrix defects, and segregation to dislocations and grain boundaries [6-8].

Current embrittlement regulations are based on correlations of surveillance data for a large number of RPV steels irradiated at low flux in operating reactors[2]. However, current $\Delta T$ models are valid only up to about one half the peak neutron fluence that would be experienced by some vessels in the PWR fleet under extended life operation, which is about $10^{24}$ n·m$^{-2}$ at 80 years (for a flux of $\approx 5 \times 10^{14}$ n·m$^{-2}$s$^{-1}$). Notably, current regulatory models under-predict test reactor $\Delta T$ for these extended-life fluences in high flux test reactor experiments [9].

It has been proposed that this under prediction is due to the emergence of slowly developing hardening and embrittlement mechanisms at high fluence [9]. The most important mechanism is nucleation and growth of Mn-Ni-Si precipitates (MNSPs), as long ago predicted by Odette and co-workers [10-13]. This early work extended previous studies by Odette that demonstrated and modeled hardening and embrittlement mechanisms associated with the rapid formation of nm-scale coherent Cu precipitates that act as obstacles to dislocation glide [14]. Researchers soon recognized that the copper rich precipitates (CRPs) are highly alloyed with Mn and Ni [10, 11]. Mn-Ni synergisms rationalized the strong effects of the alloying element Ni, as well as Cu, on hardening and embrittlement [12]. In the early 1990s Odette and coworkers carried out CALPHAD [10] based thermodynamic calculations, suggesting that Mn-Ni precipitates could form even in low Cu steels, which would otherwise be relatively insensitive to embrittlement. The Mn-Ni precipitates were predicted to be slow to nucleate and grow, thus they were dubbed late blooming phases (LBP) at that time.

---

[2] RPVs are basically one-of-a-kind individually built structures, with a wide range of steel metallurgical variables (composition, start of life microstructure and product form) that are exposed to a range of irradiation variables (neutron flux, fluence, spectrum and temperature). Embrittlement is controlled by synergistic interactions between these variables acting in combination.

These models equilibrated Mn, Ni and Cu in solution with these solutes in a specified number density of CRPs and included the effects of composition on the interface energy [10]. At sufficiently high Ni and at lower temperatures, the precipitates contain more Mn and Ni than Cu. This mean-field thermodynamic modeling was later extended to include Si in Lattice Monte Carlo (LMC) simulations, based on pair-bond energy estimates extracted from CALPHAD, that predicted Cu-rich core and Mn-Ni-Si-rich shell precipitate structures [12]. Such core shell structures were also observed in atom probe tomography studies [15-18]. The early models predicted that low irradiation temperatures, high Ni and even small amounts of Cu enhance the formation of Mn-Ni and MNSPs. It was also envisioned that MNSPs heterogeneously nucleate on small Cu-Mn-Ni-Si-defect cluster complexes that form in displacement cascades. These complexes are referred to as stable matrix features (SMF). The SMF, that are thought to be precursors to MNSPs, are responsible for low-to-intermediate fluence embrittlement in low Cu steels (< 0.07 wt.%). Since there is much more Mn + Ni + Si (typically > 2.5 at.%) in RPV steels than Cu (effectively less than 0.25 at%), large mole fractions ($f$) of MNSPs can produce correspondingly large amounts of hardening and embrittlement, that roughly scale with $\sqrt{f}$.

The early models led to a systematic search for MNSPs, especially as part of the UCSB IVAR program [19]. Further, Auger and Pareige et al. [20-22] observed dilute solute-defect clusters (Cu, Ni, Mn, Si, P, …) in French RPV steels at fluence as low as $2.5\times10^{23}m^{-2}$. As noted above, Odette has long argued that these clusters are the likely precursors to well-formed precipitates [23]. An important advance, reported by Odette in 2004, was the first observation MNSPs in a Cu-free split-melt RPV steel [24]. However, MNSPs are now widely observed in test reactor [15, 17, 18, 25, 26] and surveillance [27, 28] irradiations. It is no longer a question of if MNSPs exist, but rather to what extent they will develop as a function of flux, fluence, temperature, and alloy composition. Notably, however, MNSPs are not accounted for in current regulatory models [29].

Some aspects of the formation of these MNSPs are still under debate. In particular, the issues include: 1) whether these MNSPs are irradiation *induced* or *enhanced*; 2) whether these precipitates should be called *late* blooming phases (LBP). Regarding the first issue, simulations [30, 31] and experiments [32-37] show the segregation of Mn, Ni, and Si elements to sinks, including both defect clusters produced by cascades and network dislocations. For example, Meslin [38] observed a high number density of Mn-rich clusters in an under-saturated Fe-1at.%Mn alloy after ion irradiation. These results seem to indicate the Mn clusters are irradiation *induced*, and driven by radiation induced segregation (RIS), especially the known strong enhancement of Mn transport via interstitials [39, 40]. However, it is important to point out that RIS is highly rate dependent, due to the balance of solute fluxes to and from the sink. Thus RIS clusters that form under ion irradiation may not be present in lower dose rate neutron conditions. Further, defect accumulation would occur at the precipitates if vacancies and self-interstitial atoms (SIA) do not accumulate at the MNSPs in equal numbers, leading to observable loops or nanovoids at these sites. Such features are not observed in association with large mole fractions ($f$), average sizes ($\bar{r}$) and number densities ($N$) of MNSPs.

In support of the MNSPs being radiation *enhanced*, we note that a large volume fraction of MNSPs is indeed predicted at thermodynamic equilibrium by a CALPHAD database [41], and identified as G and $\Gamma_2$ diffracting phases in x-ray scattering studies at high fluence [42]. Moreover, these precipitates also survive long term post-irradiation annealing at 425°C, which is much higher than the irradiation temperature around 290°C, where they are obviously much more stable [43]. These factors indicate the large volume fractions of precipitates are thermally stable phases and are irradiation *enhanced*.

Nevertheless, the fact that the large volume fractions of precipitates found at high fluence are enhanced does not mean RIS is unimportant, or that associated radiation induced mechanisms are not active. Due to the small driving force to form MNSPs, segregation to sinks is likely to play an important role in nucleation stage, especially in low and intermediate solute content alloys. Messina [44] recently proposed that dislocation loops generated in cascade might act as nucleation sites for Mn-Ni precipitates. The key hypothesis is that solute segregation at loops reduces their mobility, with the result that the loop number density is similar to those of the MNSPs. However, Messina's work does not treat either segregation or solute clustering directly, so the model does not predict the observed sizes and mole fractions of MNSPs. Logically, even if RIS plays an important role in MNSP nucleation, it does not mean that they are not thermally stable phases in intermediate and high solute alloys. On the other hand, the fact that the MNSPs are stable phases at high fluence in many alloys does not mean they cannot be nucleated by RIS mechanisms, particularly in lower solute alloys.

Regarding the second issue, the term LBP is used to highlight the role of MNSPs in comparison to Cu precipitates, which form and saturate at very low fluence compared to MNSPs. The scaling factor for indexing precipitation on a fluence or dpa scale ranges from 6 to more than 50 times less for CRPs than MNSPs. Thus the term LBP does not mean that MNSPs precursors are not formed at low fluences (order $10^{23}$n-m$^{-2}$), where they are observed in the form of cascade solute cluster complexes. Rather the argument is they only form large volume fraction at high fluences compared to Cu [18, 26].

Thus the model described here builds on earlier studies of the initial formation mechanism of MNSPs [30, 31, 44, 45]. Here the objective is to develop a rigorous physical model for MNSP nucleation, growth and coarsening in low Cu RPV steels up to high fluence based on the underlying thermodynamics and kinetics. The model can be used to better understand the factors controlling MNSPs evolution. It is structured to include the essential physical mechanisms, albeit with varying degrees of approximation, while minimizing the number of fitting parameters, and is both informed by, and compared to, experiments. The model predicts the $N$, size distribution, $\bar{r}$ and $f$ of the MNSPs. The model is described in detail in Sec. 2 and SI Sec. B-G, and we briefly summarize its key characteristics here.

A cluster dynamics (CD) master equation model of the evolution of the size distribution, $N$, $\bar{r}$ and mole fraction $f$ of MNSPs as a function of fluence, flux, temperature and alloy composition has been developed. The thermodynamic terms in the master equation coefficients are extracted from a CALPHAD database. The model assumes that the MNSPs have the equilibrium compositions of either G ($Ni_{16}Si_7Mn_6$) or $\Gamma_2$ ($Mn(Ni,Si)_2$) phases. These are the two stable phases in the Mn-Ni-Si ternary projection of the Fe-Mn-Ni-Si quaternary system that are predicted by the CALPHAD database for the compositions studied in this work at around 300°C.

The kinetic terms in the CD model are based on radiation-enhanced adjustments of the estimated thermal solute diffusion coefficients, as determined by conventional reaction diffusion equations and semi-empirical flux scaling. Heterogeneous MNSP formation on the cascade cluster complexes, as noted above, is included adding to homogeneous nucleation. The results of the CD model are compared to detailed atom probe tomography (APT) data. The interface energies, heterogeneous nucleation site size and cascade cluster production efficiency factor are the main fitting parameters.

The model can be used to interpolate and extrapolate the experimental database on RPV steels, which is both sparse and not well distributed in the high dimensional space of potentially relevant variables. The CD model also provides a foundation for more realistic but complex models, e.g., including Cu and MNSPs, and simpler, but still physical, reduced order models like Avrami-based treatments of the MNSPs evolution as a function of fluence.

The paper is organized as follows. Section 2 describes the computational models and sub models. Section 3 presents the APT database. Section 4 compares the CD model predictions to the APT results. Section 5 applies the model to predict the effects of selected variations in alloy composition and irradiation temperature on the fluence dependence of the evolution of MNSPs. Section 6 analyzes the corresponding effects of composition and temperature on MNSP mole fraction at specified fluences and temperatures in terms of composition cross-plots for the entire CD database. Section 7 illustrates the effects of MNSPs on hardening and embrittlement for typical RPV steels during life-extension based on existing microstructure-property-property models. A brief summary and conclusions are presented in Section 8.

## 2. Computational Methods

### 2.1. Cluster Dynamics

Cluster dynamics (CD) can be used to describe the nucleation, growth and coarsening of precipitates from supersaturated solid solutions. The CD model used in this work is based on Slezov's general model of the kinetics of first order phase transformations [46-49], which describes the precipitate size distribution as a function of time in terms of the transition rates into ($J_{n-1 \to n}$) and out of ($J_{n \to n+1}$) cluster size $n$ from and to adjoining clusters in cluster size space. Our model also adds a heterogeneous nucleation rate ($R_{het}$) of clusters of size $n_{het}$. Figure 1 schematically illustrates the CD model. This so-called master equation for changes in the number of clusters in size class $n$ as a function of time ($f(n,t)$, m$^{-3}$) can be expressed as

$$\frac{\partial f(n,t)}{\partial t} = R_{het}(n,t) + J_{n-1 \to n} - J_{n \to n+1} \tag{1}$$

$$J_{n \to n+1} = \omega_{n,n+1}^{(+)} f(n,t) - \omega_{n+1,n}^{(-)} f(n+1,t) \tag{2}$$

The coefficients $\omega_{n,n+1}^{(+)}$ are the rates at which clusters of size $n$ absorb single atoms to grow to size $n+1$. Similarly, the coefficients $\omega_{n+1,n}^{(-)}$ are the rates at which clusters of size $n+1$ emit single particles to shrink to size $n$. The parameters needed in the homogeneous nucleation model are: a) a composition weighted effective radiation enhanced diffusion coefficient ($D_{eff}$); b) the bulk free energy of formation of a cluster phase ($\Delta g$); and, c) the cluster interfacial energy ($\gamma$). Detailed expressions used in applying this model are given in SI Sec. B and a corresponding derivation is found in Chapter 2 of Ref. [49]. More details regarding the $R_{het}$ are presented in Sec. 2.2 and SI Sec. D.

The coupled set of ordinary differential equations (ODEs) in Eq. (1) are integrated to calculate $f_n$(t) for $n = 1$ to $n_{max}$. In the case of precipitation of a single solute species with monomer transitions, as is assumed here, the integrations are trivial, even for a large $n_{max}$.

However, in this case the clusters are composed of three species, specifically Mn, Ni and Si. Thus, in principle, it would be necessary to solve clustering ODE for these three solute dimensions, potentially involving integrating up to $O(n_{max}^3)$ equations (see SI Sec. B for derivation), which would be computationally prohibitive. To avoid this complication, it is assumed that the clusters grow and shrink by absorbing or emitting pseudo-monomer molecules that are composed of Mn, Ni and Si with the stoichiometric composition of the precipitating phase [49]. There certainly can be some errors caused by this treatment, but as mentioned above, it would be computationally prohibitive with current approaches to treat all three elements individually. Further, it is observed that the compositions of the MNSPs are nearly stoichiometric and do not change much with fluence [18].

The cluster pseudo solute emission rates (defined in SI Sec. B) are based on a detailed balance between adjoining cluster sizes, which depends on the cluster free energies of formation and the radiation enhanced diffusion kinetics. The formation free energy of clusters with *n* atoms for a dilute solution model can be written in terms of a solute product as

$$\Delta G(n) = n\Delta g + \sigma(n) = nk_B T ln\left(\frac{\overline{K_{sp}}}{K_{sp}}\right) + 4\pi r_n^2 \gamma \qquad (\,3\,)$$

where $\Delta g$ is the formation free energy per atom of an infinite bulk precipitate phase, $\sigma(n)$ is the interfacial energy for clusters with *n* atoms with radius $r_n$ and an interfacial energy per unit area of $\gamma$, $k_B$ is the Boltzman constant, $T$ is temperature. The solute product, $K_{sp}$, is defined as

$$K_{sp} = \prod_i c_i^{x_i} \qquad (\,4\,)$$

Here $c_i$ is the instantaneous composition of each dissolved element in the alloy matrix and $x_i$ is the composition of the element in the precipitate phase. The solute product at equilibrium is denoted as $\overline{K_{sp}}$. The classical derivation and justification of the relation between solute products and formation energy of precipitates is provided in SI Sec. B. Equilibrium solute products are determined from CALPHAD thermodynamic models, as listed in Table 1. Interfacial energies were fit to experimental data, and are given in Table 3 (details regarding the fitting process are provided in SI Sec. G). The radiation enhanced diffusion kinetics was determined from literature values of thermal diffusion coefficients, previously parameterized models for excess vacancy radiation enhanced diffusion at low flux and a previously derived simple *effective* flux scaling model [50]. Details are provided in Sec. 2.3 and SI Sec. B and C. The code is available online at https://github.com/uw-cmg/MNS_CD, and access is available upon request prior to public release.

It is useful to draw a simple picture of what the CD model treats. Initially, precipitation is dominated by clustering and precipitate nucleation and the initial stages of growth. However, nucleation rates decrease rapidly as the solutes are depleted from the supersaturated matrix. This gives rise to an overlapping region dominated by precipitate growth. However, growth rates also decrease with dissolved solute depletion, and becomes negligible as the Gibbs-Thomson effect modified solubility limit is approached. This condition gives rise to a third overlapping region dominated by precipitate coarsening. Notably, the CD model treats all of these processes seamlessly, and in a way that directly connects to the underlying thermodynamics and kinetics.

## 2.2. Heterogeneous Nucleation

Heterogeneous nucleation is a critical element of the model. Here we assume that clusters of size $n$ are generated in displacement cascades at a rate $R_{het}(n,t)$, which has units of number of clusters generated per unit volume and per unit time. The underlying mechanism is that the high concentrations of point defects generated in the cascades, during both formation and much longer time aging periods, lead to defect-solute cluster complex formation due to corresponding defect-solute binding energies and local RIS. The first model for solute-defect complex formation (Cu-coated nanovoids) was first proposed by Odette [51] and later confirmed by a number of Kinetic Lattice Monte Carlo models of cascade aging [52-54]. While formation on both vacancy and interstitial clusters (nanovoids and loops, respectively) is possible, recent Kinetic Monte Carlo simulations suggest that MNSPs primarily form on loops created in the cascades [30, 44]. Further discussion of the rich physics of cascade aging in alloys is beyond the scope of this paper, but it is sufficient to note that large solute cluster complexes, and their solute remnants, likely form in cascades at some frequency that decreases with decreasing size and the corresponding primary knock-on atom recoil energy. Note heterogeneous nucleation also takes place on network dislocations, in association with significant solute segregation. However, this is not modeled here.

$R_{het}$ is proportional to the cascade production rate per atom, $\sigma_{cas}\phi/\Omega$, where $\sigma_{cas}$ is the cascade production cross section, $\phi$ is the neutron flux, and $\Omega$ is the atomic volume. $R_{het}$ also depends on the amount of the remaining dissolved solutes, which we represent by the ratio of the instantaneous reaction triple product $K_{sp}$ to a reference $K_{sp}^0$. Finally, the fact that at a given $K_{sp}$ only a fraction of cascades produce a large cluster of defined size $n$, is represented by a cascade cluster production efficiency factor, $\alpha$, at $K_{sp}$. While there is a distribution of $n$ that can contribute to heterogeneous nucleation, for simplicity assume that only clusters of size $n_{het}$ (fitting parameter) are generated. Thus $R_{het}$ is given by

$$R_{het}(n_{het}, t) = \alpha \cdot \frac{\sigma_{cas}\phi}{\Omega} \cdot \frac{K_{sp}(t)}{K_{sp}^0},$$ ( 5 )

and $R_{het}(n \neq n_{het}, t) = 0$.

The efficiency factor, $\alpha$, is a fitting parameter representing the fraction of cascades that produce a cluster of size $n_{het}$ when the solute product is at equilibrium ($K_{sp}(t) = K_{sp}^0$). The physical basis for $\alpha$ is partly stochastic, and partly associated with the energy of the primary knock-on atom (PKA) that creates the cascade (note, PKAs are generated by fast neutrons over a wide range of recoil energies). However, given the complexity of forming solute-defect complexes in displacement cascades, $\alpha$ is treated as the second heterogeneous fitting parameter.

While rooted in plausible physics, this nucleation model is a surrogate for more complex mechanisms of heterogeneous nucleation that also may include solute segregation to ex-cascade small loops and nanovoids. As noted above, this treatment does not include the role of dislocations and grain boundaries in heterogenoeus nucleation, although both are observed experimentally, especially for conditions of low solute supersaturations. Further, the present model does not treat the potential role of Cu. While the alloys studied here have low Cu contents, it has long been observed and predicted that small amounts of this element, even below the concentration for well-formed precipitates, can enhance MNSP nucleation [10]. Such effects, to the extent they are significant, are effectively renormalized into the fitting parameters in the model. These approximations will be dealt with in future work and are discussed further in SI Sec. D.

## 2.3. Radiation Enhanced Diffusion

The cluster dynamics model used in this work requires Mn, Ni, and Si diffusion coefficients under irradiation. The diffusion of these species is accelerated by excess vacancies produced under irradiation, a phenomenon generally referred to as radiation enhanced diffusion (RED). A model for RED is used in this work that accounts for flux effects with a simple power law-scaling model. We use a simplified version of a RED model developed by Odette et al. that accounts for solute-vacancy-trap enhanced recombination and extra annihilation of vacancies and self-interstitial atoms at transient sinks formed in cascades [50, 55, 56].

Here the thermal diffusion coefficients $D^{th}$ are given by the standard expression

$$D^{th} = D_0^{th} \exp\left(-\frac{Q}{kT}\right) \hspace{4cm} (\,6\,)$$

Unfortunately, there is essentially no diffusion data for Mn, Ni, and Si in Fe at the relevant LWR temperatures of about 300°C. Thus estimating $D^{th}$ requires a large extrapolation from high temperature. Furthermore, the low temperature $D^{th}$ must account for the effects of changing ferromagnetic state on the migration energetics of the solute below the Curie temperature. Although self-diffusion coefficients of Fe have been studied by a number of researchers, for example [57-63], there is very limited experimental data on the thermal diffusion coefficients of Mn, Ni, and Si solutes in ferromagnetic Fe. Indeed, to our knowledge, there is only one dataset for Mn [64], one for Ni [65], and none for Si. Although there are calculation results of diffusion coefficients at low temperature from DFT [66, 67], we chose to use experimental data to avoid possible uncertainties in the DFT models. In both cases small errors in migration energies can cause large differences of diffusion coefficients at the relevant low temperatures.

Therefore, we have developed an approximate, but internally consistent, approach to estimate together the diffusion coefficient for Fe and all the solutes based on constant activation energy models fit to data for the ferromagnetic host Fe matrix. Details about how the thermal diffusion coefficients were obtained are given in SI Sec. F. In the CD model the thermal diffusion coefficients for the individual solutes were then adjusted to give the RED coefficients ($D^*$, as described below) and then further combined into one effective value based on the composition of the matrix and precipitates, as described in SI Sec. C and Ref. [55].

The RED coefficient, $D^*$, can most simply be expressed as [55]

$$D^* \approx D_v X_v \frac{D^{th}}{D^{sd}} + D^{th} \hspace{3cm} (\,7\,)$$

$D^*$ is the diffusion coefficient for a given species used in the cluster dynamics model (see SI Sec. B), $D^{th}$ is the thermal diffusion coefficient of solute, $D^{sd}$ is the self-diffusion coefficient of the solvent Fe, $D_v$ is the diffusion coefficient of vacancies and $X_v$ is vacancy concentration under irradiation. At steady state, when defect production and annihilation rates balance, $X_v$ is given by [55]

$$D_v X_v = \frac{g_s \xi \sigma_{dpa} \phi}{Z_t} \qquad (8)$$

Here, $\xi$ is the cascade surviving defect production to dpa efficiency factor, $\sigma_{dpa}$ is the dpa cross-section[3], $Z_t$ is the total sink strength and $g_s(\phi, T, Z_t, ....)$ is the fraction of migrating defects that escape recombination to reach sinks [55]. The detailed model for $g_s$ accounts for flux dependent solute-trap enhanced recombination between vacancies and interstitials, with $g_s < 1$ at high flux, and $g_s \approx 1$ at low flux, when sinks dominate. The expression for $g_s(\phi, T, Z_t, ....)$ is complex and contains a number of parameters, like the vacancy trapping energy, solute trap concentrations, and the sink strength, where the latter may evolve with irradiation. Thus we have chosen a much simpler expedient to calculate $g_s$ as a function of flux that has been empirically highly successful [50, 55, 68]. This simple model approximates $g_s(\phi)$ as

$$g_s(\phi) = g_s(\phi_r)(\phi_r/\phi)^p \qquad (9)$$

Here, $\phi_r$ is a chosen low reference flux, taken as $3 \times 10^{15} \mathrm{m}^{-2}\mathrm{s}^{-1}$, and $p$ is an empirical scaling exponent. Further, $g_s(\phi_r) \approx 1$ at low flux (negligible recombination), hence, $g_s(\phi) = (\phi_r/\phi)^p$ and together with Eq. (7) and Eq. (8) $D^*$ can be expressed as a function of flux as [55]

$$D^* = D^{th} \frac{\xi \sigma_{dpa}(\phi_r/\phi)^p}{Z_t D^{sd}} \phi = k(\phi, ...)\phi \qquad (10)$$

Given the approximate estimates of $D^{th}$ and the additional complexity of RED treated by a simple model, there is no reason to expect that the $D^*$ would be accurate within the very precise constraints that would be needed for perfect agreement of the CD model with the data. However, if we allow $D^*$ to vary so as to fit the experimental mole fraction data points exactly, the resulting values fall within a factor of two of the standard fixed CD values for all alloys except Ginna (where the difference is only a factor of three), as can be seen from Figure SI-8. Finally, we note that there may be some evolution of the sink density at high fluence. However, there is insufficient information to treat this in the model. The most extreme estimates of the dislocation loop density at high fluence might double the initial network value. However, the effects of increases in the sink density at high flux are reduced by corresponding decreases in recombination rates.

---

[3] The dpa rate is the neutron flux $\phi$ (with units of neutrons/area-time) times $\sigma_{dpa}$ (with units of area/neutron-atom) times $\xi$. $X_v D_v Z_t$ is the annihilation rate of defects at sinks in the absence of recombination. The $g_s$ term accounts for defects lost to recombination.

The effective time evolution of the model is set by both the diffusion rates and the homogeneous and heterogeneous nucleation rates. Ignoring the nucleation rate effects for the moment, the scaling in Eq. (10) means that the time evolution of the system is controlled by an effective time $t_e = t(\phi_r/\phi)^p$. This time scaling allows us to define an effective fluence ($\phi t_e$),

$$\phi t_e = \phi t(\phi_r/\phi)^p \qquad (11)$$

Here $\phi t$ is the actual fluence. The $\phi t_e$ accounts for how flux affects the excess irradiation induced vacancies that accelerate diffusion. In the $g_s = 1$ sink-dominated limit there is no flux effect on $D^*t$, while in the recombination dominated limit $g_s$ is proportional to $(\phi_r/\phi)^{0.5}$. Thus higher flux delays MNSP (and CRP) evolution. Independent analysis has shown $p$ typically varies between $\approx 0.15$ and $0.35$ over a wide range of higher flux (or equivalent dpa rates for charged particle irradiations) [56]. A typical value of $p = 0.2$ was used in this work, but not adjusted to match the present data.

This very simple model for the effect of flux on RED, however, does not account for flux effects on heterogeneous nucleation rates. In particular, in our model the heterogeneous nucleation term, $R_{het}$, depends directly on flux (as described in Sec. 2.2), therefore reflects the flux independent number of cascades at a given fluence, that scales with $\phi t$ rather than with $\phi t_e$. The CD model directly accounts for both homogeneous and heterogeneous effects on nucleation, and the $p$-flux scaling approach is only used to account for radiation enhanced diffusion. We also note that the $k(\phi)$ in Eq. (10) is almost certain to vary somewhat from alloy to alloy, and for different irradiation conditions, in a way that is not fully modeled by the multi element $D^{th}$ and $p$-flux scaling approach [55]. Thus the predictions of the precipitate size distribution, $r$, $N$ and $f$ would be expected to move to higher or lower positions on the fluence scale depending on the individual case. While it could be argued that each particular set of alloy-irradiation conditions (say CM6 and ATR-1 – see below) should be individually fit for $D^*$, we chose to use only one calculated $D^*(\phi_r)$ along with the $p$-scaling flux adjustment, so that the model can be applied across many alloys and used to model new situations where fitting $D^*$ is not possible. In spite of its approximate nature and simplicity, as shown below, the unfitted treatment of $D^*$ works remarkably well within the context of the current objectives of the model. A corollary is that a much higher $D^*$, due to diffusion via SIA, is not consistent with the experimental observations.

## 2.4. Input Parameters

The thermodynamic, thermal diffusion and other parameters used in the model are summarized in Tables 1, 2 and 3, respectively. The equilibrium solute products ($\overline{K_{sp}}$) of the two phases (T3 & T6 ) at different temperatures corresponding to the nominal experimental conditions were obtained from the TCAL2 database [69] (see more in Sec. 4.1), and are listed in Table 1. Table 1 also shows the solute product at 320°C which is a newly adjusted as-run temperature for the ATR-1 data. This temperature adjustment only affects the ATR-1 data and the consequences of this recent adjustment to the corresponding predictions are discussed below.

Table 2 lists the thermal diffusion coefficient ($D^{th}$) parameters as discussed in Sec. 2.3 and further in SI Sec. F.

Table 1 The equilibrium solute product ($\overline{K_{sp}}$) for the G-phase (T3) and $\Gamma_2$-phase (T6) at different temperatures.

| Temperature (°C) | Equilibrium solute product (×10⁻³) | |
|---|---|---|
| | T3 | T6 |
| 280 | 1.96 | 2.33 |
| 284 | 2.12 | 2.53 |
| 290 | 2.21 | 2.56 |
| 300 | 2.45 | 2.82 |
| 320 | 3.02 | 3.42 |

Table 2 The thermal diffusion coefficients used in this study.

| Element | $D_0$ (×10⁻⁴ m²·s⁻¹) | Q (kJ·mol⁻¹) | Reference |
|---|---|---|---|
| Mn | 1.49 | 234.0 | [64] |
| Ni | 1.40 | 245.6 | [65] |
| Si | 0.78 | 231.5 | [70] |
| Fe | 27.5 | 254.0 | [57] |

All other parameters used in the model are listed in Table 3. Most of them were obtained from two papers [55, 71]. Four of them, denoted with FIT in parentheses, are fitting parameters, including the heterogeneous nucleation size and cascade cluster production efficiency factor (see Sec. 2.2) and two interfacial energies. Please note ATR-1 data were not used to obtain the fitting parameters, due to their irradiation condition with high flux, fluence, and irradiation temperature, although comparisons between CD results and experimental data are still given. The fitting process is discussed in detail in SI Sec. G. Note that the numerical fit values are all physically reasonable. In particular, the nucleation size $n_{het}$ is relatively small, consistent with it being formed during cascade aging. Similarly, $\alpha$ is also quite reasonable, consistent with just a small fraction of cascades (~1%) producing heterogeneous nucleation sites at low supersaturations, and is a small fraction of the 5-7 interstitial loops produced by cascades [44]. Both the T3 and T6 interfacial energies (0.185 and 0.175 J/m², respectively) are quite similar, consistent with their similar compositions. Both values are significantly lower than the Cu-Fe interfacial energy (typically taken to be about 0.4 J·m$^{-2}$ [10]), which is consistent with experimental observations that Mn, Ni and Si are usually segregated to the CRPs interface [15-18]. Finally, we note that because of our simple power law based scaling approach, $D^*$ is not sensitive to the vacancy migration energy; further details are provided in SI Sec. C.

Table 3 The model parameters used in calculating the radiation enhanced diffusion coefficient and other parameters used in cluster dynamics model.

| | |
|---|---|
| SIA – vacancy recombination radius ($r_v$, nm) | 0.57 [55]* |
| Fraction of vacancies and SIA created per dpa ($\xi$) | 0.4 [55] |
| Displacement-per-atom (dpa) cross-section ($\sigma_{dpa}$, m$^2$) | 1.5×10$^{-25}$ [55] |
| Atomic volume ($\Omega_a$, m$^3$) | 1.18×10$^{-29}$ |
| Vacancy diffusion coefficient pre-exponential factor ($D_v$, m$^2$·s$^{-1}$) | 1×10$^{-4}$ [71]* |
| Vacancy migration energy ($E_v^m$, eV) | 1.3 [71]* |
| Dislocation sink strength (dislocation density) ($\rho$, m$^{-2}$) | 2×10$^{14}$ [55] |
| Reference flux ($\phi_r$, m$^{-2}$s$^{-1}$) | 3×10$^{15}$ [55] |
| Flux effect scaling exponential factor ($p$) | 0.2 [56, 72] |
| Cascade cross section ($\sigma_{cas}$, m$^2$) | 2×10$^{-28}$ [72] |
| Reference solute product ($K_{sp}^0$) | 2.4×10$^{-3}$ |
| Heterogeneous nucleation size ($n_{het}$) (FIT) | 80 |
| Cascade cluster production efficiency factor ($\alpha$) (FIT) | 4.8×10$^{-3}$ |
| Interfacial energy of T3 phase ($\gamma_{T3}$, J·m$^{-2}$) (FIT) | 0.185 |
| Interfacial energy of T6 phase ($\gamma_{T6}$, J·m$^{-2}$) (FIT) | 0.175 |

*These values are used in calculating $g_s(\phi_r)$. $g_s(\phi_r)$ is very close to one and largely insensitive to the exact values used.

## 2.5. Size Cutoffs Used for Data Presentation and Analysis

The CD model predicts the precipitate number density $N$, size distribution $N(r)$, mean radius $\bar{r}$, and mole fraction $f$ (see SI Sec. E for definitions of how each of these are determined), as a function of fluence. The CD calculations were carried out as a function of alloy composition and temperature. As discussed below in Sec. 3, the models used the actual local chemistries of the various atom probe tips and the estimated irradiation temperature for the comparisons with experimental data. In making these comparisons, we exclude clusters below a certain size in the CD calculations, to be consistent with the minimum number of precipitate atoms in APT experiments, which is 12-30 [18]. Based on a detection efficiency of about 37% APT measurements do not detect solute clusters smaller than about 32-80 atoms. The cutoff size is 65 atoms for ATR-1, BR2-TU CM6 and Ringhals N180, 32 atoms for BR2-TU LG, BR2-G1and BR2-G2, and 43 atoms for other Ringhals and Ginna alloys, respectively. The cutoff size used for analysis of composition and temperature effects, where no direct comparison to experiment is presently being made, is simply set to be 32 atoms, which is the minimum size used in the APT analysis. Finally, we exclude clusters sizes below 40 atoms (~0.49nm in radius) for all mechanical property calculations in Sec. 7. The cutoff size for mechanical property calculations is chosen based on that recommended in the Russell-Brown model [73]. We note that no cutoffs are used in the actual CD microstructure simulations, and are only used when presenting results. It is shown in Figure SI-7 that there is almost no effect on the mole fraction of precipitates from these small clusters, so they are expected to have little impact on the mechanical properties of the alloy.

## 3.   Atom Probe Tomography Data Used to Validate and Calibrate the CD Model

Five RPV steels irradiated in four reactors were studied. The irradiations included the high flux test reactors, Advanced Test Reactor (ATR) and Belgian Reactor 2 (BR2), and low flux reactor surveillance capsules from the Ringhals and Ginna power plants. Flux and fluence varied from $5.9 \times 10^{14}$n·m$^{-2}$s$^{-1}$ to $2.3 \times 10^{18}$n·m$^{-2}$s$^{-1}$ and $3.3 \times 10^{23}$n·m$^{-2}$ to $1.1 \times 10^{25}$n·m$^{-2}$ (or dose from 0.0495 to 1.65dpa [55]), respectively. The nominal temperature ranged from 284°C to 300°C. However, a recent as-run reevaluation placed the ATR-1 temperature at ≈320°C, so this adjusted value was used in this work. The test reactor irradiations included Cu free plate (LG) and forging (CM6) that contain ≈ 0.75 to 0.85 wt% Ni and 1.35 to 1.65 wt.% Ni, respectively. These steels have bainitic microstructures and properties characteristic of in-service steels. The steels in the surveillance irradiations were all low Cu (< 0.07 wt.%) steels with intermediate (Ginna) and high (Ringhals) Ni contents, respectively. The data sets are identified by the irradiation condition and alloy designation. However, since the main variables are the Ni content, flux and fluence, in addition to the identification cited above, a parenthetical notation is added to designate a low (L), medium (M) or high (H) category, for these compositional and irradiation variables, respectively. The parenthesis gives a 3-tuple whose order corresponds to Ni content, flux, and fluence. For example, Ringhals N is categorized as (HLM), meaning high Ni, low flux and medium fluence, while ATR-1 LG as (MHH), meaning medium Ni, high flux and high fluence. Note that these designations do not account for variations in alloy Mn and Si contents.

Table 4 The alloy compositions and irradiation conditions in this study.

| Name (Ni, flux, fluence) | Composition (at.%) | | | Irradiation Condition | | | Reference |
|---|---|---|---|---|---|---|---|
| | Mn | Ni | Si | Temperature (°C) | Flux (×$10^{16}$ n·m$^{-2}$s$^{-1}$) | Fluence (×$10^{23}$ n·m$^{-2}$) | |
| Ringhals E (HLM) | 1.25 | 1.50 | 0.41 | 284 | 0.147 to 0.166 | 6.39 | [27], this work |
| Ringhals N (HLM) | 1.14 | 1.68 | 0.28 | | | 3.3, 6.03 | |
| Ginna Forging (MLM) | 0.57 | 0.92 | 0.67 | 290 | 0.0592 | 5.80 | [28], this work |
| BR2-TU LG (MMH) | 1.18 | 0.73 | 0.44 | 300 | 30 | 25 | This work |
| BR2-TU CM6 (HMH) | 1.16 | 1.63 | 0.35 | | | | |
| BR2-G2 CM6 (HHM) | 1.07 | 1.46 | 0.34 | 300 | 100 | 6.67 | This work |
| BR2-G1 LG (MHH) | 1.09 | 0.86 | 0.49 | 300 | 100 | 13.3 | [18] |
| BR2-G1 CM6 (HHH) | 1.09 | 1.34 | 0.33 | | | | |
| ATR-1 LG (MHH) | 0.87 | 0.71 | 0.43 | 320 | 230 | 110 | [18] |
| ATR-1 CM6 (HHH) | 1.42 | 1.69 | 0.39 | | | | |

It was previously shown that nano-scale MNSPs respond to local alloy compositions of individual APT tips [18]. The database used in this study included up to nine APT tips per alloy, yielding a total of 36 different compositions. The average APT measured compositions and irradiation conditions are shown in Table 4. The full local compositions, when more than one tip was measured, are reported in supplementary information Table SI-1. The nominal Cameca

Integrated Visualization and Analysis (IVAS) precipitate compositions, including the indicated Fe in all tips, are provided in Table SI-2.

The APT analysis method used for all the data obtained by UCSB is described in detail in reference [18]. Previously published APT on Ringhals welds [27] and Ginna forging [28] were reanalyzed by Edmondson in a way that is consistent with [18] to provide a maximal consistent data set.

## 4. Results

### 4.1. Fe-Mn-Ni-Si System Thermodynamics at ≈ 300°C

The Thermo-Calc [74] Aluminum 2 (TCAL2) CALPHAD database [69] was used to establish the Fe-Mn-Ni-Si phase diagram as the thermodynamic foundation for this study. This thermodynamic analysis updates previous work of Xiong et al. [41]. It has been recently determined that the ATR-1 data shown in [41] was actually run at 320° rather than 290°C which was previously believed based on thermal models due to higher than initially assumed heating rates. Thus an update of the thermodynamic calculation results using a temperature of 320°C, as well as an explicit discussion of the Gibbs-Thomson effect on the mole fraction of the MNSPs is given in SI Sec. H.

The TCAL2 database predicts that there are two intermetallic phases at equilibrium at around 300°C: the T3 or G-phase, ($Mn_6Ni_{16}Si_7$); and the T6 or $\Gamma_2$ phase, ($Mn(Ni,Si)_2$). Their respective crystal structures are shown in Table 5 [75, 76]. Figure 2a shows that the predicted precipitate compositions compared to experimental results, for the average APT tip compositions and irradiation temperatures in this study, are in good general agreement (with RMS of 5.17%, 7.54% and 6.13% for Ni, Mn and Si, respectively). The corresponding predicted equilibrium mole fractions ($f$) of each element in the MNSPs are shown in Figure 2b. No comparison is made with experiment in this case, since the nm-scale MNSPs are not in equilibrium even at the highest fluence due to a significant Gibbs-Thomson effect (comparisons are made in SI. Sec. H for the ATR-1 condition). Figure 2c shows the predicted MNSPs and their mole fractions for each alloy. Based on CALPHAD calculation results, in 6 out of the 10 cases the precipitates are the stoichiometric T6/$\Gamma_2$ phase. The T6 phase is generally associated with high Ni alloys (>1.6wt. %); however, one high Ni alloy had a very small amount of T3/G phase. One of the intermediate Ni alloys contains stoichiometric T6 phase, one other contains stoichiometric T3 phase, while two others contain both T3 and T6 phases; one has about 62% T3 and 38% T6, and the other has about 12% T3 and 88% T6. The mole factions of these phases decrease with increasing temperature. For a nominal alloy composition of 1.0 Mn-0.6Si-0.75Ni at.%, the T3 phase persists up to 408°C, while T6 phase persists up to 430°C. When Ni is increased to 1.6 at% the T3 phase persists up to 455°C and T6 phase persists up to 487°C. Note all results mentioned above are based on CALPHAD calculations. Details aside, the major conclusion is that equilibrium MNSP intermetallic phases can readily form in the ≈ 270 to 300°C operating temperature regime of RPVs. This result is in contradiction with a number of first-principles and kinetic Monte Carlo studies arguing that MNSPs must be irradiation induced [30, 31, 77].

Table 5 The crystal structures of Mn-Ni-Si phases in RPV steels at around 300°C.

| Phase | Composition | Space group | Type | Lattice constant, nm | Number of atoms in a unit cell |
|---|---|---|---|---|---|
| T3/G-phase | $Mn_6Ni_{16}Si_7$ | $Fm\bar{3}m$ | $Mg_6Cu_{16}Si_7$ | 1.108 | 116 |
| T6/$\Gamma_2$ | $Mn(Ni,Si)_2$ | $Fd\bar{3}m$ | $Cu_2Mg$ | 0.668 | 24 |

### 4.2. Radiation Enhanced Precipitation Kinetics

The CD model was used to predict the evolution of MNSPs as a function of fluence, temperature, flux and alloy composition. Figure 3 and Figure 4 show the evolution of MNSPs ($N$, $\bar{r}$ and $f$) as a function of fluence for all the alloys. The bands represent the range of outputs obtained by simulating individual APT tip compositions. Note that the CD model predicts that the number density and mole fraction of the lower Ni, higher temperature ATR-1 LG (MHH coding) MNSPs are so low that they are not visible on scales used in Figure 3. The corresponding evolutions of MNSPs for the average APT tip compositions are shown in Figure SI-4. Comparisons of the CD model precipitate size distributions between and the experimental data for selective alloys are shown in Figure SI-5. The CD model predicts all the alloys are in the nucleation and growth (N&G) stages of precipitation except in the case of the high Ni ATR-1 CM6 (HHH) condition, where the early precipitation and very high fluence of $1.1 \times 10^{25} \, \text{m}^{-2}$ extends into the coarsening regime. Note in the N&G regime the model-experiment comparisons are very sensitive to the accuracy of the simplified RED treatment of $D^*$, as discussed in Sec. 2.3.

Figure 5 shows the strong effect of higher flux on delaying precipitation by comparing CD model predictions for nearly identical alloys (Ringhals N and CM6), both with high $\approx 1.6$ at. % Ni, irradiated at widely different fluxes ($1.49 \times 10^{15} \, \text{m}^{-2}\text{s}^{-1}$ for Ringhals and up to $2.3 \times 10^{18} \, \text{m}^{-2}\text{s}^{-1}$ for ATR-1). The BR2-TU flux ($3 \times 10^{17} \, \text{m}^{-2}\text{s}^{-1}$) was chosen as the reference condition for the corresponding CM6 data (filled squares). Thus the higher flux CM6 data (unfilled circles and diamonds) for the ATR-1, G1, G2 conditions, shown in Table 4, are also plotted as filled symbols that have been adjusted, as indicated by the arrows, to an effective fluence for the BR2-TU reference flux, using the simple $p = 0.2$ scaling model in Eq. (9). The solid line shows the CD predictions for the Ringhals N average composition and the nominal low flux irradiation condition at $1.49 \times 10^{15} \, \text{m}^{-2}\text{s}^{-1}$ and 284°C. The shaded band shows the corresponding CD results for the CM6 average composition at the BR2-TU reference flux for irradiations at 300 and 320°C, covering the range of experimental temperatures.

Notably, the experimentally observed effect of flux is somewhat larger than predicted by the CD model, which can be seen from the fact that the difference in fluence for an approximately fixed precipitate volume fraction is larger in the experiments than in the CD model. It is useful to be reminded that, as discussed in Sec. 2.3, the fluence scale of the two curves is largely governed by $D^*$, and that besides being oversimplified, the use of the canonical average $p = 0.2$ increases the approximate nature of the $D^*$ model. For example, it is known that the higher Ni contents in CM6 and Ringhals lead to a higher rate of solute-vacancy-trap enhanced recombination, which in turn corresponds to a larger effective $p$ [55]. More generally the agreement between the CD model predictions and experiment would be enhanced if $D^*$ itself were treated as a fitting parameter for individual alloys. Figure 5 also shows that the precipitation plateau for CM6 predicted by CD at very high fluence is only $\approx 2/3$ of experimental data, reflecting the approximate nature of the thermodynamic model.

Figure 6 provides a more detailed one-on-one comparison of the predicted cluster dynamics precipitate $\bar{r}$, $N$ and $f$ with APT measurements. If we exclude the ATR-1 experiments, the agreement is reasonable across this large range of compositions and irradiation conditions. However, the mole fractions of the MNSPs during the growth stage for the medium Ni alloys (BR2-TU MMH LG, BR2-G1 MHH LG and MLM Ginna) are all under-predicted by the model. These differences are most likely caused either by uncertainties in the thermodynamic parameterization and RED model or unmodeled physics, where the latter may include dislocation and, especially, small loop enhanced nucleation and solute segregation, including that driven by radiation induced segregation (RIS). The predicted and measured values for LG (LHH) and CM6 (HHH) under the ATR-1 irradiation conditions (using the recently updated temperature of 320°C) are shown as the open symbols. These results were not included in the parameter fitting, since the very high flux, fluence and high temperature irradiation condition is well beyond the corresponding range for LWR service and more relevant test reactor data. However, it's still useful to show the comparisons between the CD model predictions and the experimental data for this extreme condition. MNSPs are significantly under-predicted by the CD model in this case, particularly for LG. Actually, the experimentally observed mole fractions of MNSPs for these alloys are even higher than what were predicted by CALPHAD model at equilibrium without considering the Gibbs-Thomson effect (see SI Sec. H). Therefore, unmodeled physics might be playing a significant role, including that which is due to the very high flux used in the ATR-1 experiment. For example, this could be related to the influence of small amounts of Cu, or segregation at small, surviving, ex-cascade dislocation loops. In general, dislocation assisted nucleation is expected, especially for cases like LG with lower Ni concentration and free energy differences driving precipitation [28]. Note after the MNSPs have grown, their small loops nucleation sites would be consumed, or not apparent, making their role difficult to identify.

Overall, except for the confounded ATR-1 data, the CD model predictions are in semi-quantitative agreement with experimental APT measurements of MNSP $N$, $\bar{r}$ and $f$, as well as the observed MNSP size distribution. Given the complexity of multicomponent precipitation under irradiation and the wide range of compositions and irradiation conditions considered, this level of agreement suggests that the model reasonably represents the dominant underlying physics of MNSPs evolution under irradiation. Most importantly, the model can be used to better understand the role of different mechanisms, and assess the influence of key material and irradiation variables for low Cu steels under both test reactor and vessel service LWR irradiation conditions.

As an example of possible mechanistic insight, Figure 7 shows the ratio of number density of MNSPs formed by heterogeneous nucleation to homogeneous nucleation at fluences of 1, 5 and $10 \times 10^{23}$ n·m$^{-2}$. In all cases heterogeneous nucleation is dominant at lower fluence and in 5 out of 8 cases it is dominant even at high fluence. Plots of MNSP number densities for different phases formed by different nucleation mechanisms as a function of fluence are shown in the supplemental information Figure SI-6.

In summary, the success of the present model suggests that vacancy enhanced diffusion is consistent with observed MNSP growth kinetics. However, other factors may also play a role, and in fact appear to be critical in the nucleation stage, where radiation induced effects associated with cascades mediate nucleation rates for medium Ni steels associated with relatively lower driving forces for homogeneous nucleation. Our results suggest that such radiation-induced processes play a dominant role in MNSP formation in low and medium Ni alloys. Furthermore, the discrepancies in the model predictions for medium Ni steels suggests that the simple model for heterogeneous nucleation used in this work may need refinement, e.g., to take into account more explicitly the role of small interstitial loops formed in cascades and segregation to these loops, as discussed in Ref. [44], as well as to pre-existing network dislocations.

## 5. Calibrated CD Model Predictions of Effects of Composition and Temperature on MNSPs

### 5.1. The Effect of Composition

Atom probe tomography (APT) is a powerful tool for three-dimensional nano-analytical mapping with near atomic-scale resolution [78]. However, the volume of the analyzed region is typically $\approx 1$ to $4 \times 10^4$ nm$^3$, corresponding to $\approx 5$ to $40 \times 10^6$ atoms and 100-200 nm in length.

Such tiny volumes are subject to significant tip-to-tip composition fluctuations in typical micro-segregated steels. For example, if a particular tip contains, or is adjacent to, a large carbide, it is likely to contain significantly less than the average Mn content, which is always lower than the nominal bulk content in any event. Both Si and Ni are controlled by the specific alloy solute additions, but these also vary somewhat from tip to tip. Thus, as has been shown [18], the precipitation in a tip is strongly affected by the local solute concentration. Indeed tip-to-tip chemistry variations can be exploited to establish dissolved solute-precipitate relationships [18]. In this work we used the actual measured APT Mn, Ni and Si compositions in modeling MNSPs evolutions. Figure 8 highlights the compositional sensitivity for irradiations at a flux of $10^{16}$ n·m$^{-2}$s$^{-1}$ at 290°C. Figure 8a shows the effects of intermediate Ni (0.80, 1.10at%) and lower Mn contents (0.80, 1.10at%), while Figure 8b is for high Ni (1.45, 1.75at.%) and slightly higher Mn (1.00, 1.30at.%). These values are in the range of the in-service type RPV alloys studied here. As expected, the MNSP $N$ and $f$ systematically increase with Mn and Ni. The corresponding $\bar{r}$ values are somewhat less sensitive to composition variations, especially at higher Ni. Composition also systematically affects both the pre- and post-saturation regimes of MNSP $N$ and $f$. Clearly, higher Ni has a powerful effect on lowering the fluence associated with the rapid MNSP formation with high $N$ and $f$. Further, in both cases, there are only slightly overlapping regions of nucleation and growth, since $N$ is nearly saturated (or decreasing slightly) when $f$ begins to rapidly increase. This observation rationalizes some of the confusion regarding the character of MNSPs. That is, a high number density of sub nm MNSPs form before they begin to grow rapidly, and are not easily identified as well-formed precipitates. Note the dominant nucleation mechanism also varies between the medium (heterogeneous) and high Ni (homogeneous) regimes. The different effects of alloy composition are discussed further in Sec.

6 below and the corresponding effects of the computed $f$ on hardening and the resulting $\Delta T$ for different alloys for low flux reactor conditions is presented in Sec. 7 and SI Sec. I.

### 5.2. Effect of Temperature

MNSPs evolution is also very sensitive to the irradiation temperature, which is significant for a number of reasons. First the temperatures in this study varied over a nominal range of $\approx 290$ to 300°C, and in practice up to $\approx 320$°C in the actual as-run condition for ATR-1. Typical in service power reactor vessel temperatures range from $\approx 270$°C to 300°C. Further, it is extremely difficult to accurately control and measure temperatures to uncertainties less than $\approx \pm 10\text{-}15$°C. Notably, the effects of temperature are also very sensitive to the alloy composition.

Figure 9 shows the effect of temperature on the evolution of MNSPs, modeled for two compositions: a high solute 1.45at.%Mn-1.65%Ni-0.45%Si and a low solute 1.00at.%Mn-0.70%Ni-0.35%Si composition, respectively. The results are shown for a flux $10^{16}\text{n}\cdot\text{m}^{-2}\text{s}^{-1}$ up to $10^{25}\text{n}\cdot\text{m}^{-2}$ at 280, 290 and 300°C. The effect of these variations of temperature mainly occurs during the nucleation and growth stage of precipitation associated with the rapid increase in $N$, which can differ up to five times for high solute alloy for only a 20°C temperature difference. The effect of temperature on $N$ is also reflected on mole fraction of MNSPs that can differ up to 1.0% for high solute alloys and 0.5% for low solute alloys. There is only a minor corresponding effect on the mean MNSP radius, $\bar{r}$. These results demonstrate that temperature effects must be carefully considered when comparing experimental with the CD model predictions.

## 6. Analysis of the Large CD Database on the MNSP Mole Fraction ($f$)

### 6.1. Alloy Composition and Fluence Effects

In this section, we model the combined effect on MNSPs mole fraction of a range of RPV alloy compositions, irradiation temperatures ($T$) and fluences ($\phi t$) presented as a series of cross plots of $\sqrt{f}$ versus a single variable, while holding the other variables constant. As discussed below, irradiation hardening ($\Delta\sigma_y$) and embrittlement ($\Delta T$) primarily depend on the $\sqrt{f}$. Figure 10 shows cross plots of the $\sqrt{f}$ versus Ni, over the specified range of Mn and Si at $10^{23}$, $5\times10^{23}$ and $10^{24}$ n·m$^{-2}$ at 290°C and a flux of $3\times10^{15}$ n·m$^{-2}$s$^{-1}$. As discussed below, the $\sqrt{f}$ is used since it's the primary MNSP characteristic that controls $\Delta\sigma_y$ and $\Delta T$.

Clearly Ni has a dominant effect on $\sqrt{f}$. The effects of Mn and Si are significant but more modest than they are for Ni. The absolute MNSP $\sqrt{f}$ is low at $10^{23}$ n·m$^{-2}$, but increases somewhat starting at $\approx 1.5$at.%Ni. The effect of Ni is much stronger at $5\times10^{23}$ n·m$^{-2}$ above $\approx 0.5$at.%Ni, and $\sqrt{f}$ again increases rapidly above $\approx 1.5$at.%Ni. At $10^{24}$ n·m$^{-2}$ $\sqrt{f}$ increases approximately linearly, or with a weak polynomial dependence, between 0.5at.% and 1.6at.%Ni, and at higher Ni the increase in $\sqrt{f}$ begins to taper off. SI Sec. J provides a simple polynomial fit for $\sqrt{f}$ as a function of alloy composition at $10^{24}$ n·m$^{-2}$ and 290°C.

### 6.2. Temperature Effects

The MNSP $f$ also depends strongly on the irradiation temperature ($T$). Figure 11 plots $\sqrt{f}$ versus $T$ at Ni = 0.6at.%, 1.0at.%, 1.4at.% for 0.6at.%Si and 1.0at.%Mn at fluences of $5\times10^{23}$ and $10\times10^{23}$ n·m$^{-2}$. Figure 11 a) shows that the absolute $\sqrt{f}$ increases with increasing Ni and decreasing $T$. Figure 11 b) shows the same data normalized to 1 at 290°C. In all cases the $\sqrt{f}$ versus $T$ follows an approximately linear relation, and overall the trends are qualitatively similar.

# 7. Yield Stress Shift ($\Delta\sigma_y$) and Ductile to Brittle Transition Temperature Shift ($\Delta T$) Based on the CD-based Precipitation Model Predictions.

In order to provide perspective on the results of this study, we applied models previously developed by Odette et al [11, 18, 50, 56] to translate CD predictions of volume fraction ($f_V$), number density ($N$) and mean radius ($\bar{r}$) to yield stress increases ($\Delta\sigma_y$) and to the associated ductile-to-brittle transition temperature shift ($\Delta T$) with irradiation (see SI Sec. I for details). Note we do not claim that these examples represent actual quantitative predictions of anticipated embrittlement behavior. Figure 12 shows the predicted time and fluence dependence of $f_V$ and $\Delta T$, respectively, for three different Ni levels (0.6at.%, 1.0% and 1.6%) with 1.00at.%Mn-0.40%Si (a medium solute concentration of Mn and a typical one for Si) at 290°C and a flux of $3.6\times10^{14}$m$^{-2}$s$^{-1}$, pertinent to RPV service. Figure 12a shows that the MNSPs are still in the rapid growth stage from 40 to 80 years. At an extended life fluence of $10^{24}$n·m$^{-2}$ typical RPV steels with 1.0at.%Ni are predicted to contain $f_V \approx 0.2\%$ MNSP that is capable of producing $\Delta T \approx 80°C$. High 1.6%Ni steels contain $f_V \approx 0.9\%$ MNSP at $10^{24}$ n·m$^{-2}$, with corresponding $\Delta T$ of 250°C. A map of the CD predictions for a wider range of compositions can be found in SI Sec. I.

## 8. Conclusions

A cluster dynamics model has been developed to study the evolution of Mn-Ni-Si precipitates (MNSPs) in low-Cu RPV steels, which lead to irradiation embrittlement. The model draws upon available thermodynamic and kinetic data, and includes a semi-empirical model for radiation-enhanced diffusion, including treatment of flux effects. The model also includes heterogeneous nucleation on damage created in displacement cascades. The interfacial energies for MNSPs and heterogeneous nucleation rates and nucleated cluster sizes were determined by fitting to a large body of high quality experimental APT data. The model, which was calibrated to 10 alloys (with 36 different APT tip compositions) for a range of irradiation conditions, can semi-quantitatively predict MNSPs evolution in RPV steels. The major deficiency in the CD model is for alloys with very high flux (ATR-1), where MNSPs are either under predicted, or do not form, in contrast to experimental observations. In part this is due to the fact that the model under-predicts MNSP formation in the alloys studied here with medium Ni contents around 0.8 wt.%, and comparison to a wider body of data not considered in this work supports that this is a general trend. We discuss various possible unmodeled physics explanations for these underpredictions that will be addressed in future research.

The principle conclusions derived from this study include:

1. A thermodynamically-based CD model has been developed to treat MNSP evolution in irradiated low Cu steels that predicts the slow precipitation of G/T3 and $\Gamma_2$/T6 phases at around 290°C, that can reach high mole fractions at high fluence.

2. The CD model is generally in semi-quantitative agreement with experimental APT observations on the MNSP compositions, $N$, $r$ and $f$.

3. A heterogeneous mechanism for nucleating MNSPs is critical to their formation in intermediate and low solute alloys, although not at the high Ni contents of greater than $\approx$1.5at.%Ni, where homogeneous nucleation dominates. Heterogeneous nucleation is not needed at sufficiently high Ni due to a larger free energy difference driving precipitation.

4. The alloy Ni content is the dominant compositional factor in forming MNSPs, while Mn and Si play lesser roles. The dominant role of Ni is due to the fact the G and $\Gamma_2$ phases respectively contain 1 and 0.8 (Mn + Si) atoms for every Ni atom, respectively.

5. The threshold for G and $\Gamma_2$ MNSP formation appears to be $\approx$ 0.5at.%Ni.

6. The model suggests that MNSPs in low Cu alloys in RPVs are in a nucleation and growth stage in present reactors, with significant potential for more hardening and associated temperature shifts expected during life-extension.

7. As an illustrative example, at an extended life fluence of $10^{24}$n·m$^{-2}$ typical RPV steels with 1.0at.%Ni and 1.6at.%Ni are predicted to contain $f_V \approx$0.2% and 0.9% MNSPs that are nominally capable of producing $\Delta T \approx$80°C and 250°C, respectively.

8. The mole fractions of MNSPs under very high flux at very high fluence (ATR-1) observed from experiments are higher than the value predicted by CALPHAD model at equilibrium even without considering Gibbs-Thomson effect. Thus even with cascade enhanced nucleation, the CD model under-predicts the mole fraction of MNSPs in this case. This suggests that unmodeled other physics, like influence of small amounts of Cu, or segregation resulting in precipitation on small dislocation loops, may play a significant role under this extreme condition where full precipitation is expected.

9. The MNSP mole fraction of medium Ni steels (BR2-TU, BR2-G1 LG and Ginna) are under predicted compared to experiments. The differences may again be due to either thermodynamic model inaccuracies, or unmodeled physics.

10. We note that a major limitation of the model for real RPVs is that Cu is not treated in the CD model, but it is well established that even small concentrations, below the amounts needed to produce well-formed Cu enriched precipitates, catalyze the formation of MNSPs.

Although Cu is not included in the present model, it will serve as a foundation for future modeling of Cu-Mn-Ni-Si precipitation in Cu bearing RPV steels. Further, many simplifications and approximations in the current CD will be improved and additional physics included, especially treating nucleation in a more rigorous fashion. Notably, the CD model, and its improved progeny, will be compared to and informed by a growing RPV steel database, especially the results of the UCSB ATR-2 high fluence irradiation program. Finally, these detailed physics-based models will guide reduced order model, but still physically based, correlation equations fit to the actual surveillance and test reactor databases. Ultimately, the more detailed models themselves might be fit these databases to provide an even more detailed physical basis for embrittlement prediction interpolation and extrapolation.

## Acknowledgements

This work has been primarily supported by the US Department of Energy Office of Nuclear Energy's Light Water Reactor Sustainability Program, Materials Aging and Degradation Pathway. Partial support for Huibin Ke was provided by the DOE Office of Nuclear Energy's Integrated Research Project (IRP) under contract DE-NE0000639. A portion of this research, both the ATR irradiation and FIB/APT at the Center for Advanced Energy Studies - Microscopy and Characterization Suite (CAES-MaCS), was supported by the Advanced Test Reactor National Scientific User Facility through the U.S. Department of Energy, Office of Nuclear Energy under DOE Idaho Operations Office Contract DE-AC07-051D14517. We also would like to thank Dr. Takuya Yamamoto at the University of California-Santa Barbara who helped with the hardening model in this work. Special thanks go to Collin Knight at the Idaho National Laboratory for assisting in the complex steps needed to gain access to the irradiated ATR specimens. The much earlier piggyback BR2 irradiations were sponsored by Jean Claude Van Duysen of Electricite de France and carried out under the supervision of Lorenzo Malerbra at SCK Belgium. We also thank Peter Hosemann at UC Berkeley for access to their FIB to prepare APT samples from the BR2 (TU, G1 and G2) conditions. The MRL Shared Experimental Facilities were used for performing APT on the BR2 samples and are supported by the MRSEC Program of the NSF under Award No. DMR 1121053 as a member of the NSF-funded Materials Research Facilities Network (www.mrfn.org).

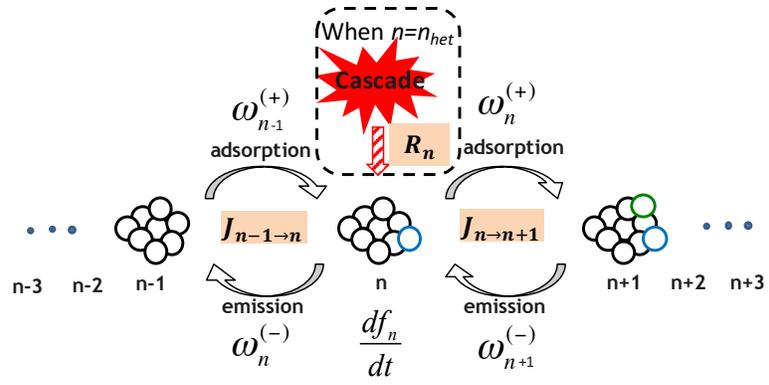

Figure 1 Schematic plot of homogeneous and heterogeneous nucleation mechanisms applied used in the cluster dynamics model.

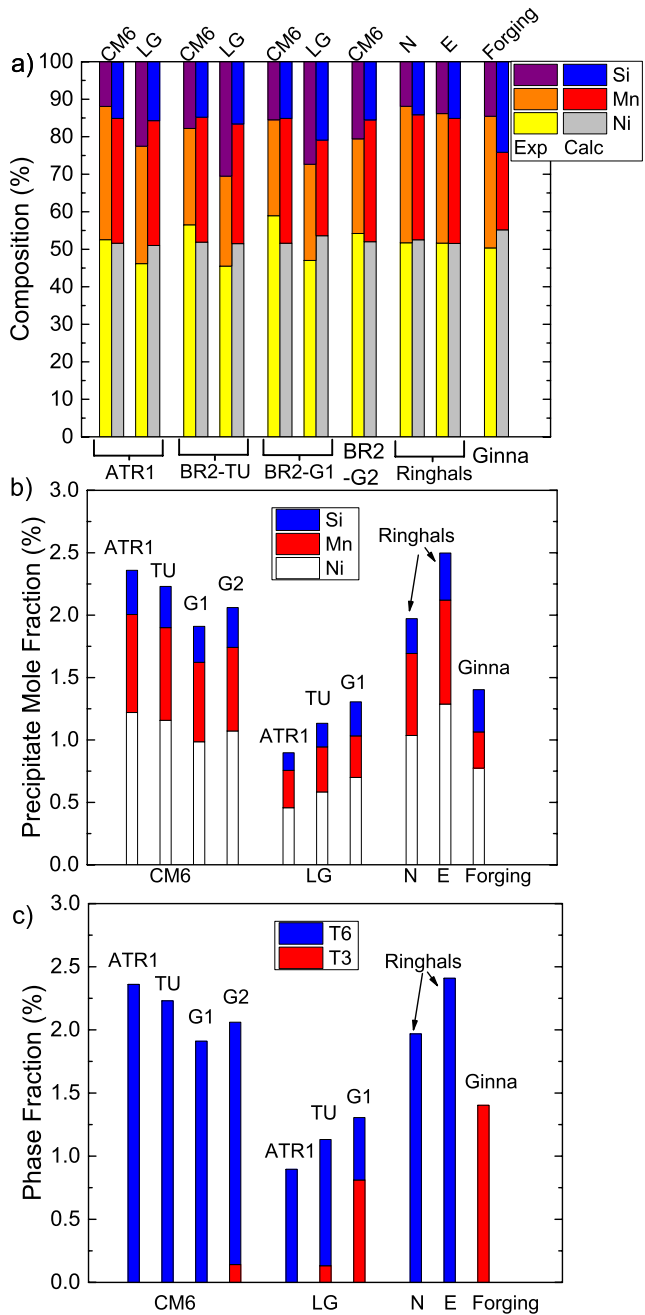

Figure 2 Thermodynamic equilibrium CALPHAD predictions.
a) precipitate compositions compared to APT data; b) precipitate solute mole fractions; c) phase selection fractions.

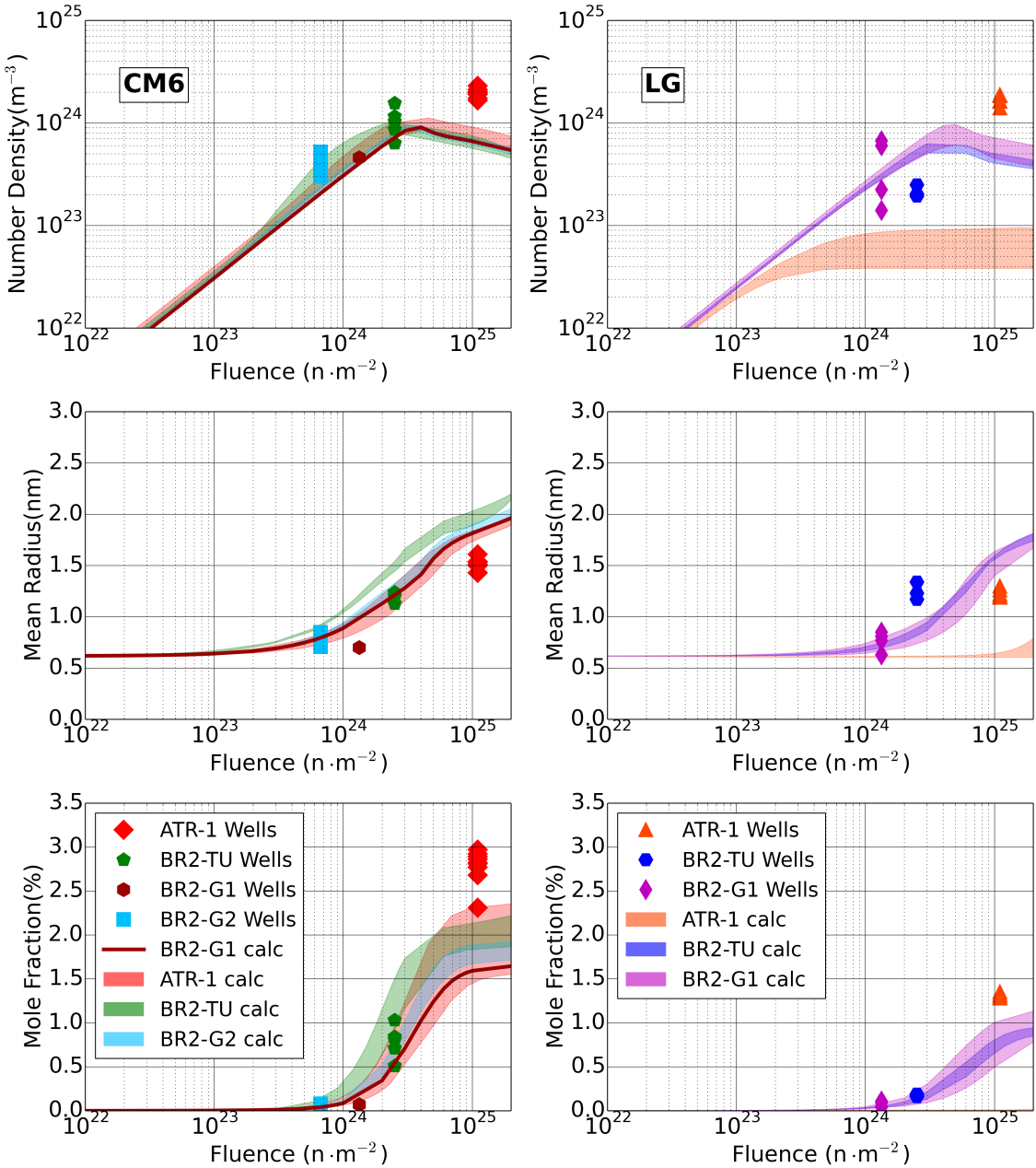

Figure 3 Evolution of precipitates in the high Ni CM6 and the medium Ni LG as a function of fluence compared to the APT data for the various irradiation conditions. The lines represent simulated data for single APT tip composition, while for multiple APT tips and associated compositions are shown as the shaded bands.

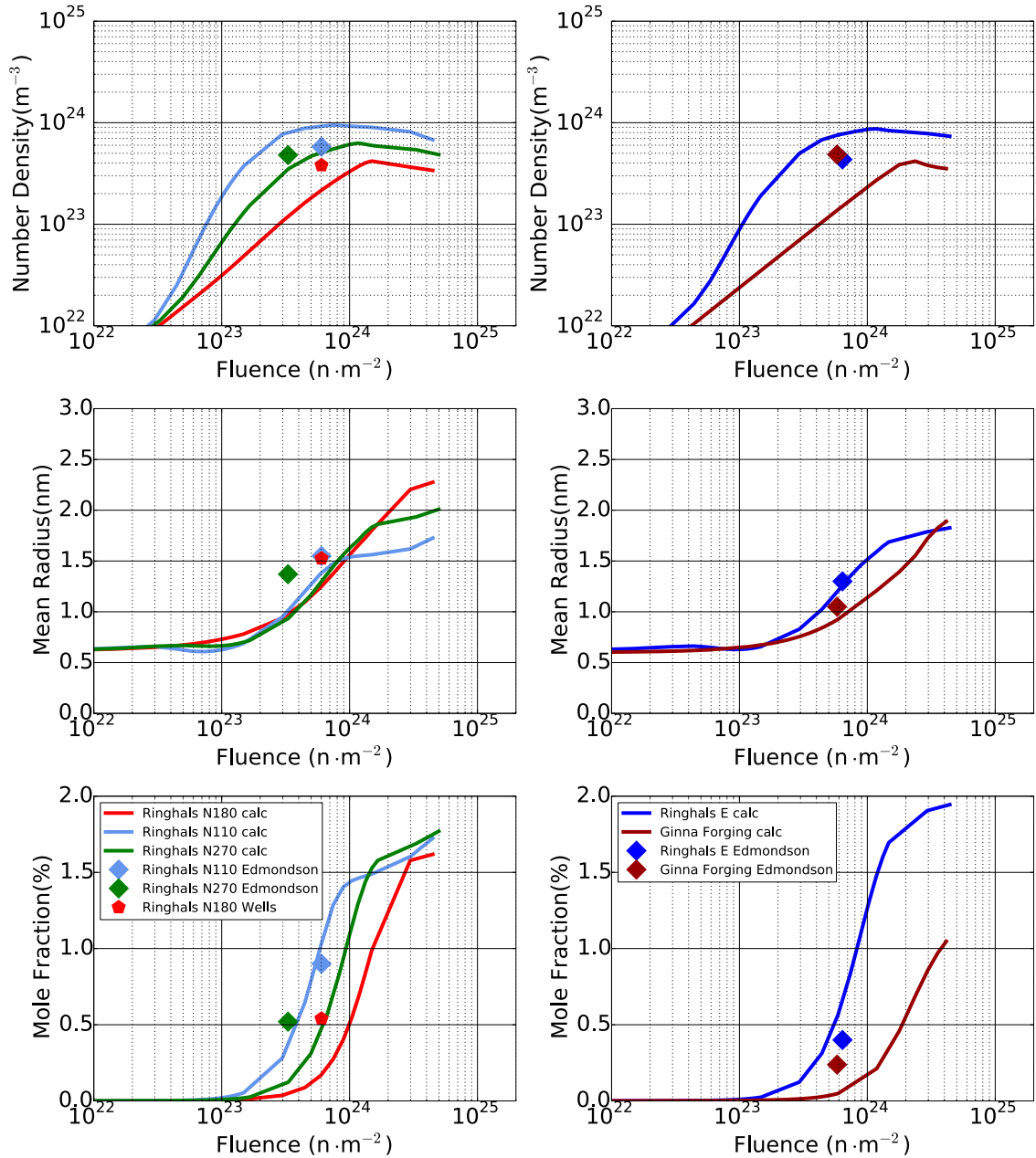

Figure 4 Evolution of precipitates as a function of fluence compared to the APT data for the various irradiation conditions for Ringhals and Ginna power reactors.

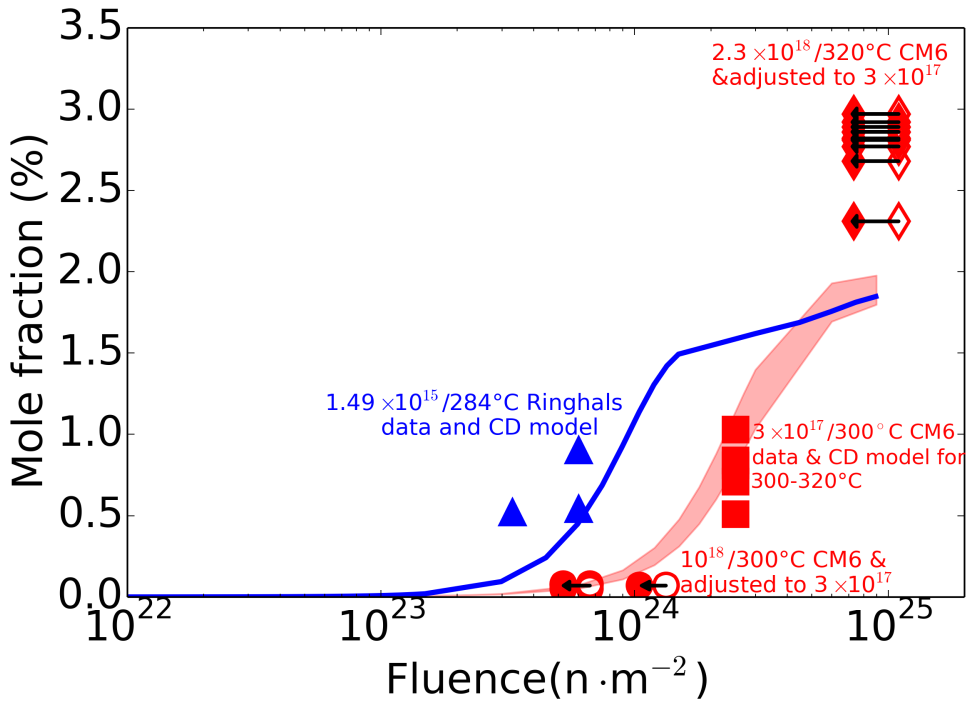

Figure 5 A comparison of the nearly identical Ringhals N alloy irradiated at low flux (blue triangles) and the CM6 alloy irradiated at high flux (filled and unfilled red circles, squares and diamonds). The open symbols are plotted at the actual fluence, while the filled circles and diamonds are plotted at an effective fluence corresponding to the BR2-TU flux, chosen as the CM6 reference condition. The BR2-TU square symbols are for the reference flux, hence are not adjusted. The blue solid line shows the CD prediction for the Ringhals N average composition and low flux irradiation condition, while the red band shows the CD predictions for the CM6 average composition at the BR2-TU reference flux for the actual irradiation temperatures ranging from 300 to 320°C.

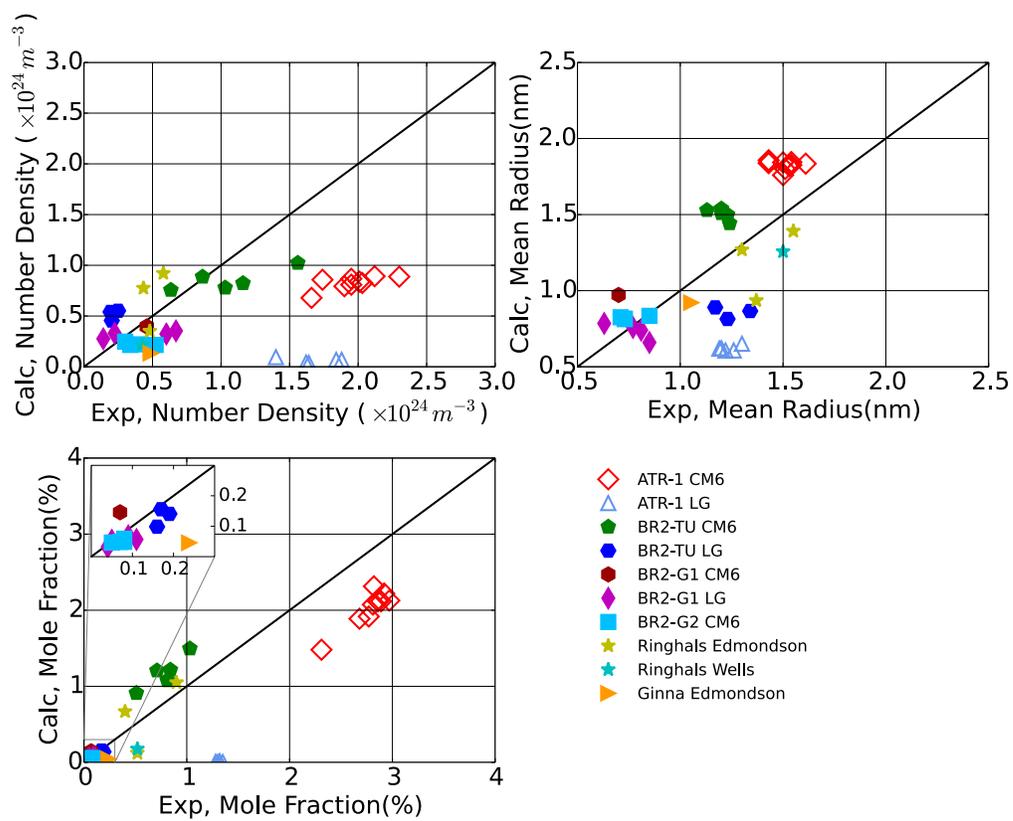

Figure 6 Comparison of the CD number density, mean radius and mole fraction and the APT data.

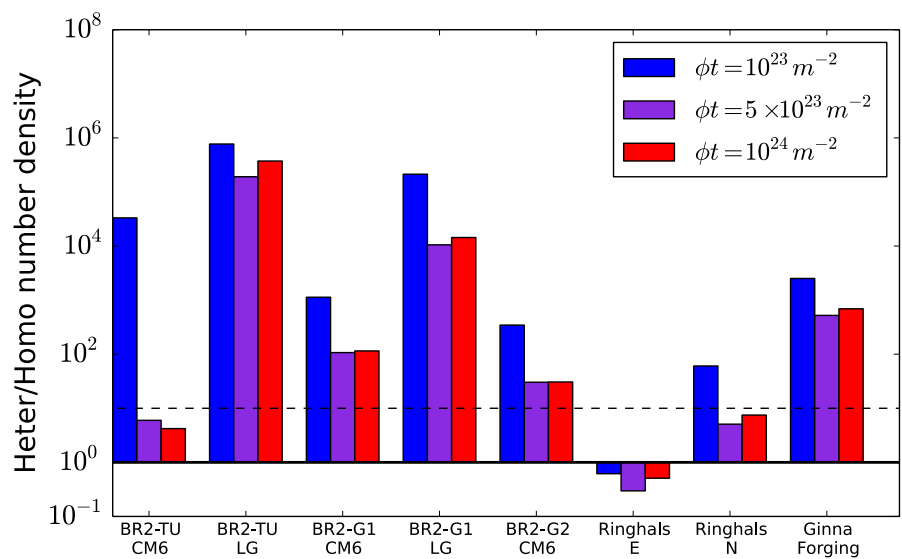

Figure 7 The ratio of number of precipitates created by heterogeneous to homogeneous nucleation at different fluences.

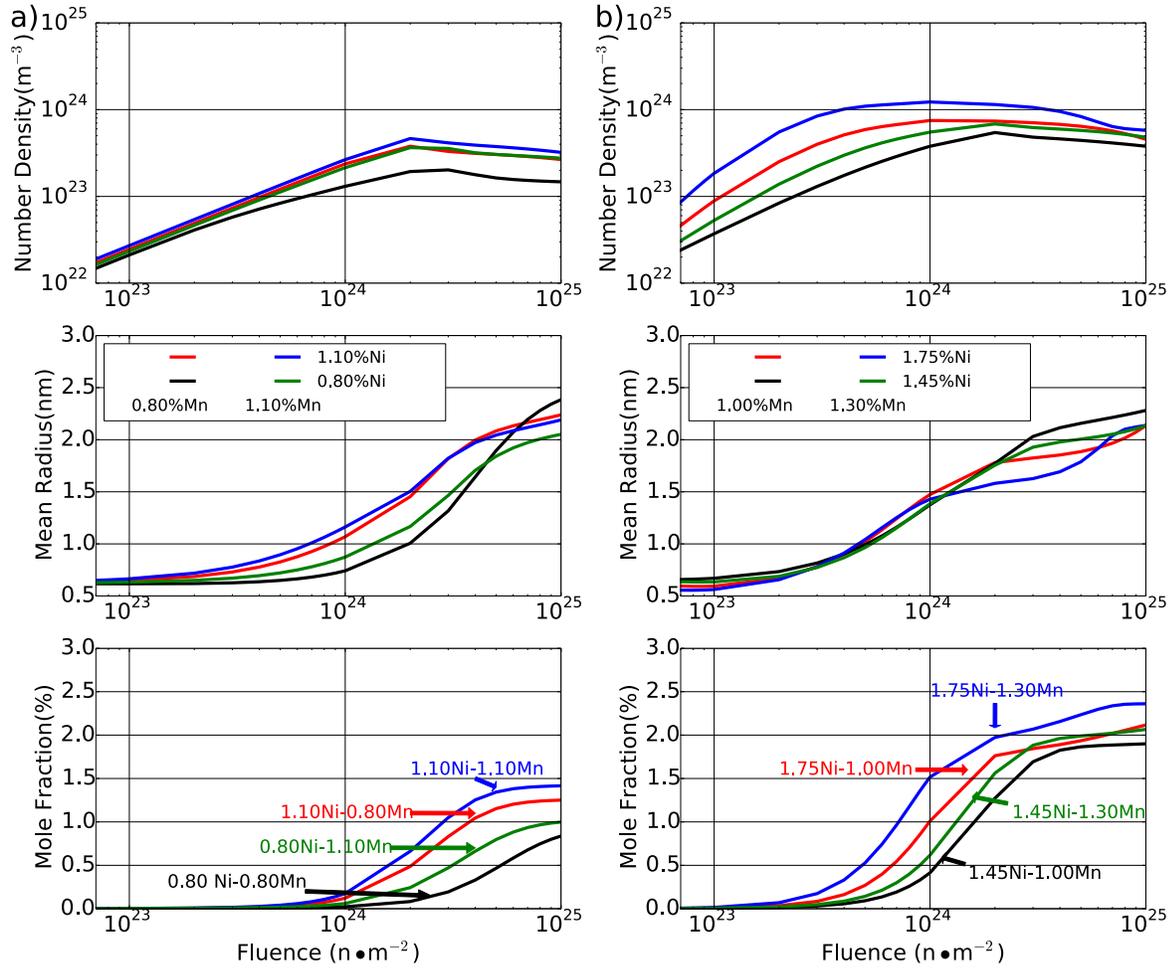

Figure 8 The effects of Mn and Ni on the evolution of precipitates at $1\times10^{16}m^{-2}s^{-1}$ at 290°C for: a) 0.35at.% Si and b) 0.45at.% Si.

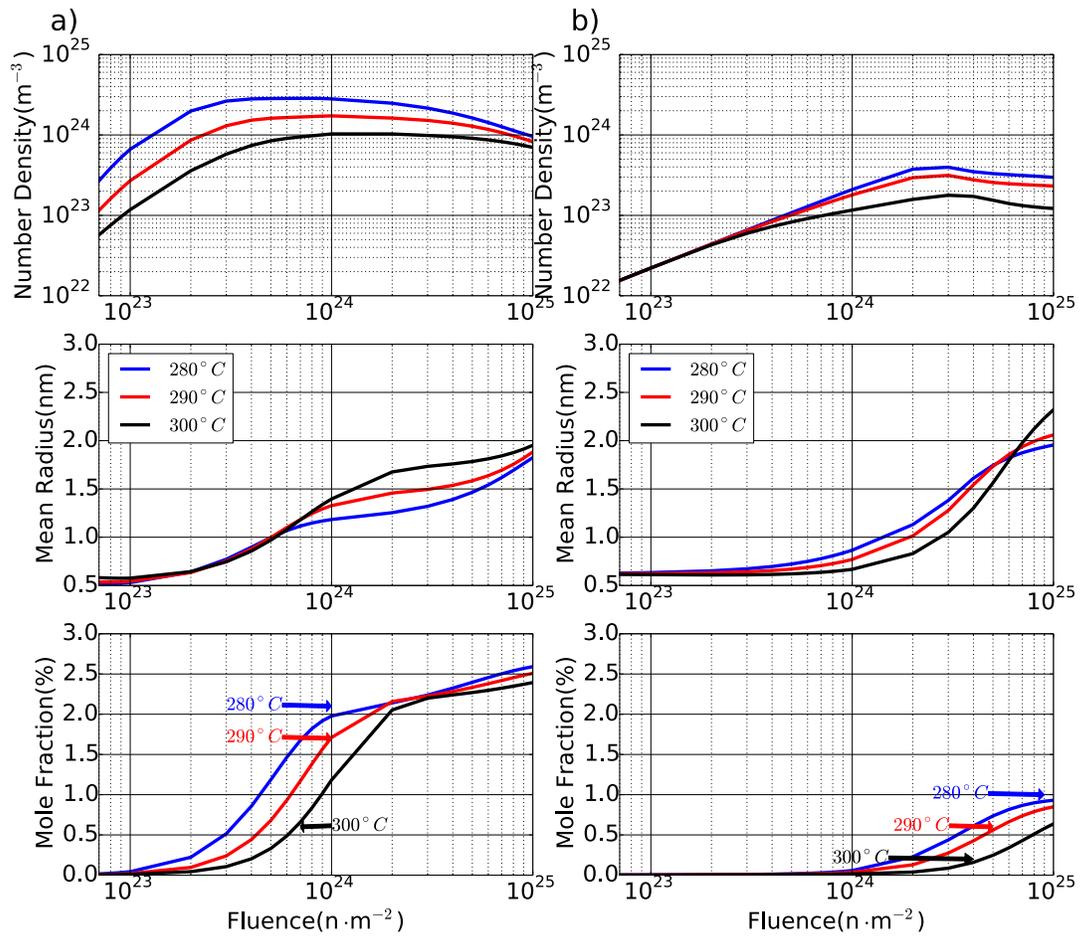

Figure 9 The effect of temperature on the evolution of MNSPs for: a) Fe-1.45at.%Mn-1.65%Ni-0.45%Si; and, b) Fe-1.00at.%Mn-0.70%Ni-0.35%Si.

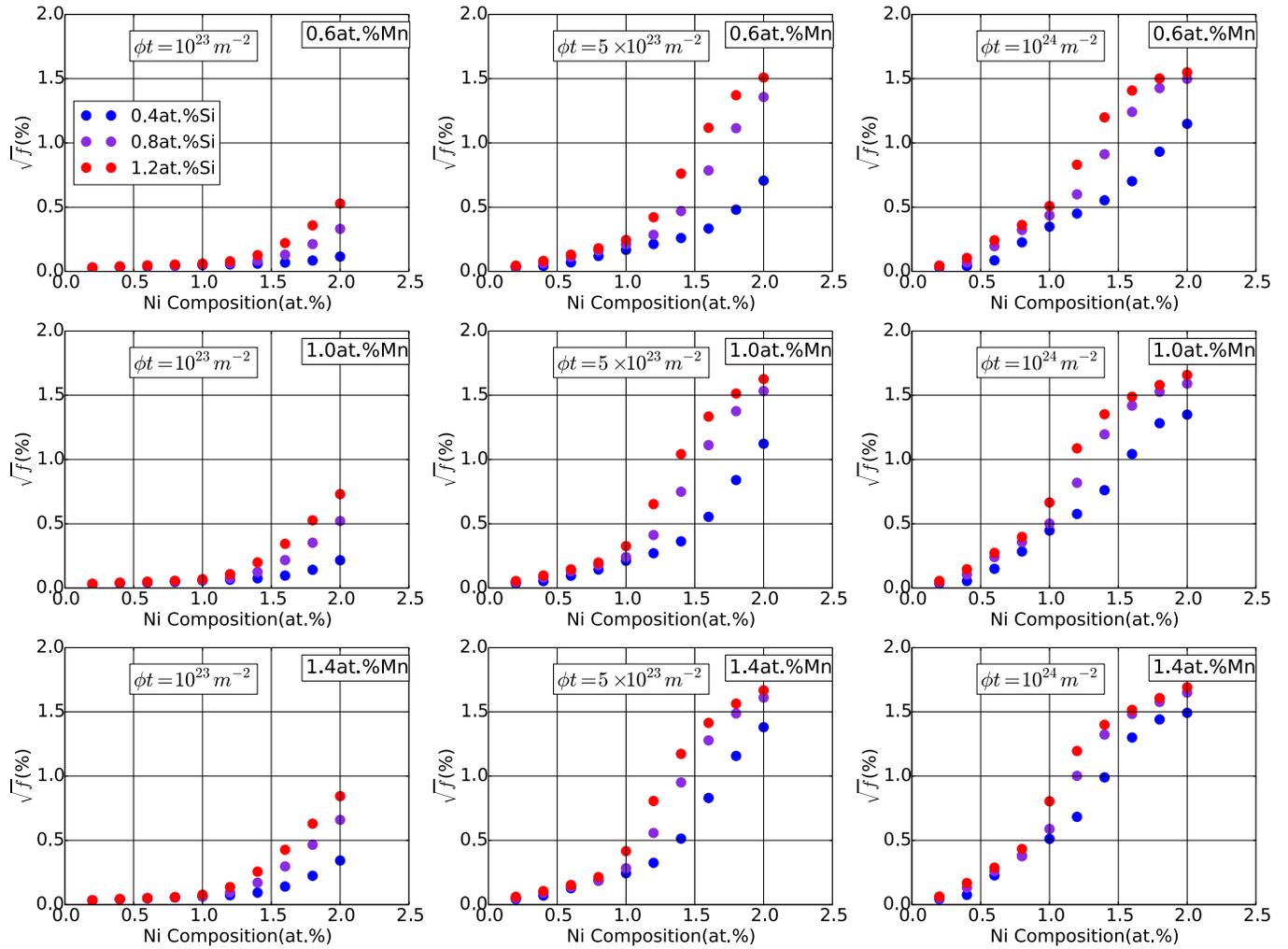

Figure 10 The square root of mole fraction ($\sqrt{f}$) as a function of Ni composition for various Mn and Si contents at different fluences.

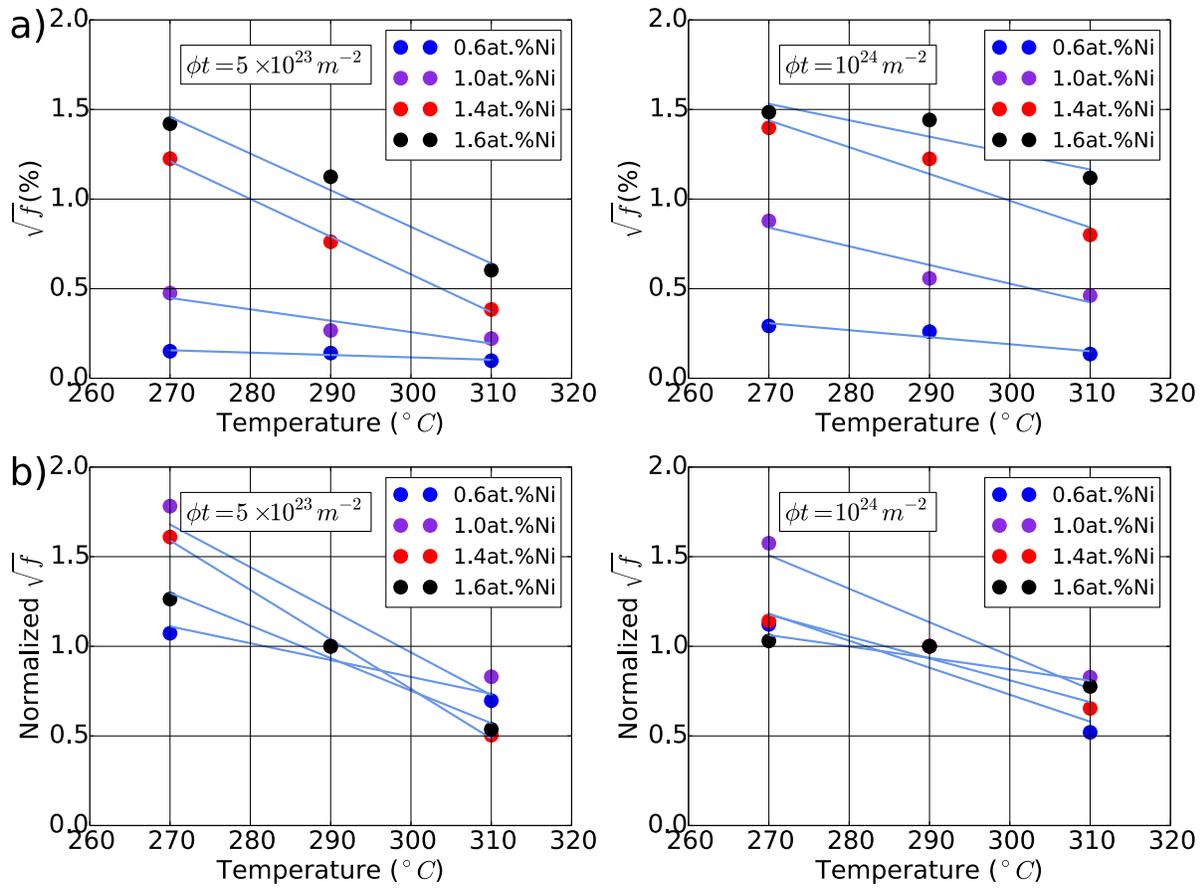

Figure 11 The effect of temperature on $\sqrt{f}$ for various alloy Ni contnets with 1.4at.%Mn-0.6%Si and two fluences: a) absolute √f; and, b) normalized to 1 at 290°C.

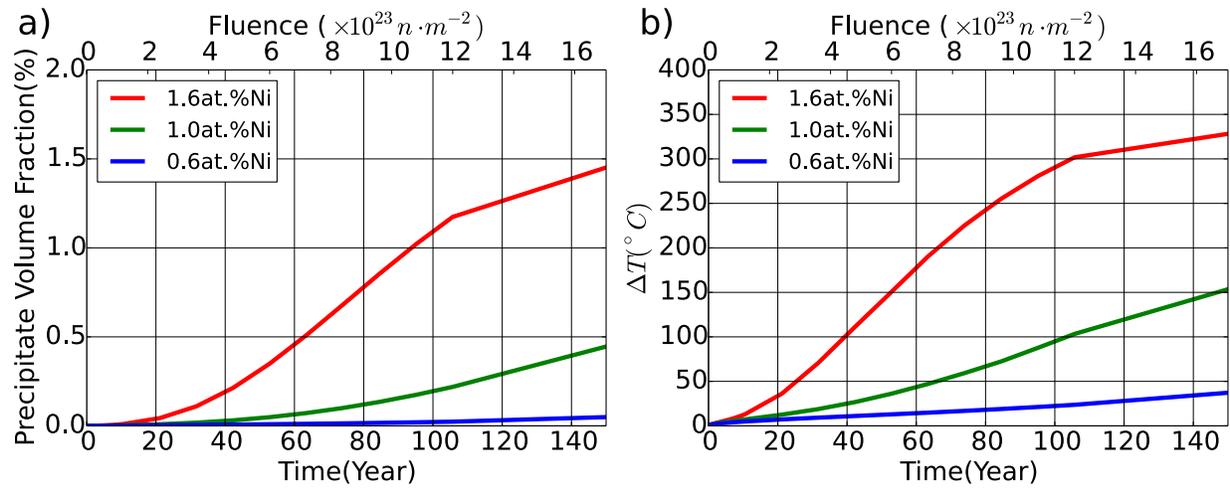

Figure 12 CD model predictions of the effect of Ni at 1.00at.%Mn-0.40%Si and a flux of $3.6 \times 10^{14}$m$^{-2}$s$^{-1}$ at 290°C: a) Volume fraction; and, b) $\Delta T$.